\documentclass[amsmath,amssymb,aps,prb,twocolumn,floatfix,10pt,superscriptaddress]{revtex4-2}

\usepackage{mathtools}
\usepackage[caption = false]{subfig}
\usepackage{graphicx}
\usepackage{standalone}
\usepackage{dcolumn}
\usepackage{bm}
\usepackage{xcolor}
\usepackage{physics}
\usepackage{enumitem}
\usepackage{calligra}
\usepackage[colorlinks = true,linkcolor = red,citecolor = magenta]{hyperref}
\usepackage[sort&compress]{natbib}
\usepackage{orcidlink}
\usepackage[normalem]{ulem}

\usepackage{pdfpages}

\makeatletter
\AtBeginDocument{\let\LS@rot\@undefined}
\makeatother

\DeclareGraphicsExtensions{.pdf,.eps,.png,.jpg,.mps}

\newcommand{\pcsadd}{Center for Theoretical Physics of Complex Systems, Institute for Basic Science (IBS), Daejeon, Korea, 34126}
\newcommand{\ustadd}{Basic Science Program, Korea University of Science and Technology (UST), Daejeon 34113, Republic of Korea}
\newcommand{\nziasadd}{Centre for Theoretical Chemistry and Physics, The New Zealand Institute for Advanced
Study (NZIAS), Massey University Albany, Auckland 0745,
New Zealand}

\makeatletter
\renewcommand*{\fnum@figure}{{\normalfont\bfseries \figurename~\thefigure}}
\renewcommand*{\@caption@fignum@sep}{\textbf{:}}
\makeatother

\newcommand{\fb}{\mathrm{FB}}

\newcommand{\db}{\mathrm{DB}}
\newcommand{\cls}{\mathrm{CLS}}
\newcommand{\bcls}{\mathrm{BCLS}}

\newcommand{\br}{\mathbf{r}}
\newcommand{\bk}{\mathbf{k}}
\newcommand{\bq}{\mathbf{q}}

\newcommand{\bz}{\textrm{BZ}}

\newcommand{\appropto}{\mathrel{\vcenter{
  \offinterlineskip\halign{\hfil$##$\cr
    \propto\cr\noalign{\kern2pt}\sim\cr\noalign{\kern-2pt}}}}}

\newlength\mylen
\newlist{mycases}{enumerate}{1}
\setlist[mycases,1]{label=\textbf{Case~\arabic*.}, 
  labelwidth=\dimexpr-\mylen-\labelsep\relax,leftmargin=0pt,align=right}

\begin{document}

\title{Real space decay of flat band projectors from compact localized states}

\author{Yeongjun Kim\,\orcidlink{}}
    \email{yeongjun.kim.04@gmail.com}
    \affiliation{\pcsadd}

\author{Sergej Flach\,\orcidlink{}}
    \email{sflach@ibs.re.kr}
    \affiliation{\pcsadd}
    \affiliation{\ustadd}
    \affiliation{\nziasadd}

\author{Alexei Andreanov\,\orcidlink{0000-0002-3033-0452}}
    \altaffiliation{Current address: Center for Trapped Ions Quantum Science, Institute for Basic Science, Daejeon 34126, Republic of Korea}
    \email{aalexei@ibs.re.kr}
    \affiliation{\pcsadd}
    \affiliation{\ustadd}

\date{\today}

\begin{abstract}
    Flatbands (FB) with compact localized eigenstates (CLS) fall into three main categories, controlled by the algebraic properties of the CLS set: orthogonal, linearly independent, linearly dependent (singular).
    A CLS parametrization allows us to continuously tune a linearly independent FB into a limiting orthogonal or a linearly dependent (singular) one.
    We derive the asymptotic real space decay of the flat band projectors for each category.
    The linearly independent FB is characterized by an exponentially decaying projector and a corresponding localization length $\xi$, all dressed by an algebraic prefactor.
    In the orthogonal limit, the localization length is \(\xi=0\), and the projector is compact.
    The singular FB limit corresponds to \(\xi \rightarrow \infty\) with an emerging power law decay of the projector.
    Furthermore, we identify an anisotropic singular flat band class with direction-dependent band touchings, which exhibits anisotropic algebraic decay of the flat band projector.
    This anisotropy is unavoidable in three-dimensional two-band lattices with real CLS.
    We obtain analytical estimates for the localization length and the algebraic power law exponents depending on the dimension of the lattice and the number of bands involved.
    Numerical results are in excellent agreement with the analytics.
    Our results are of relevance for the understanding of the details of the FB quantum metric discussed in the context of FB superconductivity, the impact of disorder, and the response to local driving.
\end{abstract}

\maketitle


\section{Introduction}
\label{sec:introduction}

Flat bands (FB) emerge in certain lattice models as bands with zero dispersion, leading to complete suppression of transport~\cite{derzhko2015strongly,leykam2018artificial,rhim2021singular,danieli2024flat}.
FB Hamiltonians with finite-range hopping support compact localized eigenstates (CLS)~\cite{read2017compactly}, eigenstates that turn strictly zero outside a finite volume due to destructive interference. 
The recent surge of interest in FBs is due to their non-trivial response to external perturbations such as
unconventional superconductivity~\cite{cao2018unconventional} and robust fractional quantum Hall effects in twisted bilayer graphene~\cite{xie2021fractional} due to nearly flat band/narrow dispersion, ferromagnetism~\cite{derzhko2015strongly, lieb1989two, mielke1991ferromagnetism, tasaki1992ferromagnetism, mielke1999stability}, anomalous 
Landau levels~\cite{rhim2020quantum}, disorder~\cite{chalker2010anderson,goda2006inverse,cadez2021metal, kim2023flat, lee2023critical, lee2023critical2}, many-body FB localization~\cite{kuno2020flat_qs, danieli2020many, vakulchyk2021heat, danieli2022many, tilleke2020nearest}, and compact discrete breathers~\cite{danieli2018compact,danieli2021nonlinear,danieli2021compact}.
Flat bands have been realized in multiple experimental settings~\cite{nakata2012observation, mukherjee2015observation, kajiwara2016observation, vicencio2015observation, nguyen2018symmetry, ma2020direct, taie2015coherent, ozawa2017interaction, baboux2016bosonic, masumoto2012exciton, wang2022observation, wang2019highly, zhou2023observation, kang2020topological, tacchi2023experimental,chasemayoral2024compact, lape2025realization}. 

FB Hamiltonians can be obtained from the knowledge of the exact CLS using FB generators which assume a CLS and find the Hamiltonian supporting this CLS as an eigenstate~\cite{maimaiti2017compact,maimaiti2019universal,maimaiti2021flatband,graf2021designing,hwang2021general}.
We classify flat band models according to their CLS set properties.
The set is obtained from translated copies of an irreducible CLS and can be (a) orthonormal (orthogonal), (b) linearly independent (but not orthogonal), and (c) linearly dependent~\cite{danieli2024flat}.
Flat bands from class (c) are also coined \textsl{singular}. 
The simplest orthogonal FBs are obtained with a CLS residing in just one unit cell~\cite{flach2014detangling}.
Linearly independent FBs are typically gapped away from a relevant dispersive part of the band structure~\cite{maimaiti2017compact,chasemayoral2024compact,di2025dipole}.
Singular FBs exist only in lattice dimension \(d\geq 2\) and necessarily band touches at least one dispersive band in at least one point in \(k\)-space~\cite{bergman2008band,rhim2019classification,rhim2021singular,graf2021designing}.
One dispersive band results in double degeneracy and quadratic dispersion close to the degeneracy point (quadratic touching), while two dispersive bands can result in triple degeneracy, be (at least locally) symmetry related and have linear dispersion close to the degeneracy point (linear touching).

The algebraic features of the CLS set  determine the real space properties of the FB projectors.
Weakly perturbed FBs, for example, in flatband superconductivity or disordered flatbands, are effectively described by projection of a perturbation onto the flat band eigenspace~\cite{cadez2021metal,lee2023critical,kim2023flat,chasemayoral2024compact,hofmann2023superconductivity}.
This operation involves the flatband projector, whose properties directly determine the properties of the resulting effective model:
within the leading order of the degenerate perturbation theory~\cite{messiah2014quantum}, a weak perturbation \(\lambda V\) of an isolated flat band produces an effective model in the flat band subspace -- \(H_\mathrm{eff}=\lambda P_{\fb} V P_{\fb}\)~\cite{cadez2021metal}.
The range and spatial structure of the effective Hamiltonian are directly controlled by the real-space decay of the flat-band projector.
The FB projector is also an experimentally relevant quantity: 
in experiments, prepared artificial lattices are often driven locally.
The response to such a driving at the FB energy is controlled by the FB projector and can be used to study superconductivity and the impact of disorder, among others~\cite{chasemayoral2024compact, di2025dipole, lape2025realization}.

\begin{figure*}[htpb]
    \centering
    \includegraphics[width=0.9\linewidth]{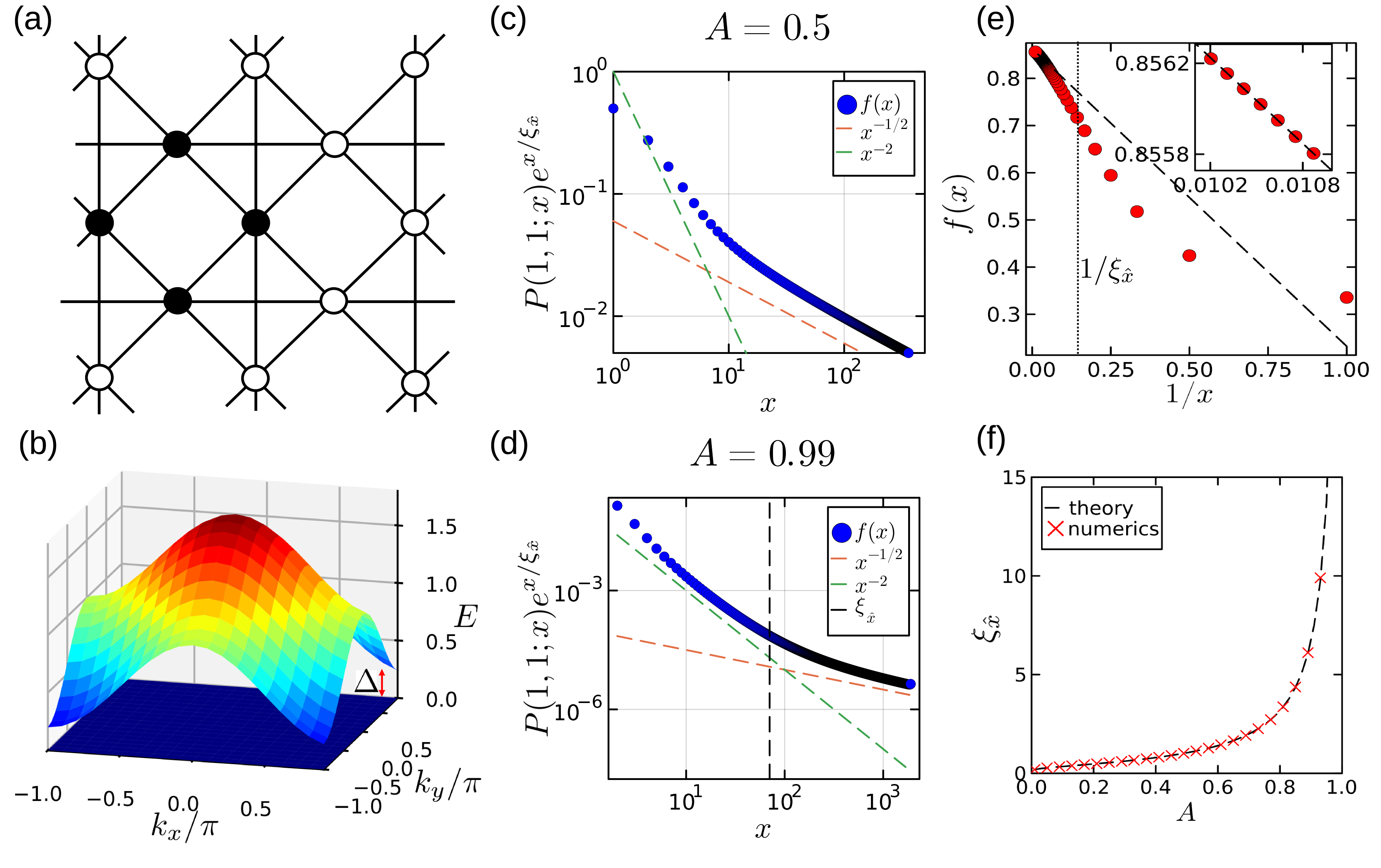}
    \caption{
        The generalized 2D checkerboard lattice. Note that the black lines represent hoppings with tunable values.
        (a): The lattice structure. 
        Black circles represent a CLS [See Eq.~\eqref{eq:square_cls}]. 
        For the values of hoppings and CLS amplitudes, see Eq.~\eqref{eq:checkerboard_hops} of Appendix.
        (b): The band structure.
        The two bands are \(E_{\fb} = 0\), and \(E_{\db}(k) = \alpha^2(k)\). 
        Here, \(A = 0.5\).
        \(\Delta\) is the band gap.
        (c) Algebraic part of the projector \(P(x)e^{x/\xi}\) shown in log-log plot.
        The red and green dashed lines guide the eye for \(x^{-1/2}\) and \(x^{-2}\) respectively.
        The parameter $A=0.5$ places the FB model halfway between orthogonal (\(A=0\)) and singular (\(A=1\)) limits.
        (d) Same as (c) but \(A=0.99\).
        This places the FB model close to the singular limit.
        The vertical black dashed line indicates the value of the localization length \(\xi_{\hat{x}} = 70\)
        (e): The ratio \(f(x)\)~\eqref{eq:ratio_test}
 vs. \(1/x\).
        Here \(A=0.9\).
        The straight black dashed line is obtained from a linear fit of the smallest \(1/x\) data (see inset).
        Its intercept value with the y-axis is $0.86$ and results in a numerical estimate of \(\xi=6.7\) in excellent agreement with our analytics.
        The slope of the dashed line is \(-0.54\) (after dividing by y-intercept) in excellent agreement with the analytical prediction \(-{\rm e}^{-1/\xi}/2\)
         (the value gets close to \(-0.5\) as larger system size increases).
        Deviations from the dashed line are observed precisely at \(1/x=1/\xi\) as predicted by analytics and indicated by the vertical dotted line.
        (f): Localization length \(\xi_{\hat{x}}\) versus \(A\).
        Symbols - numerical results from the above intercept fitting.
        Dashed line - analytics. 
    }
    \label{fig:projector_result_summary}
\end{figure*}

\section{Main results}
\label{sec:main_results}

We obtain the asymptotic real space decay of FB projectors for the above three FB classes~\cite{leykam2018artificial,rhim2021singular, danieli2024flat}:
\begin{itemize}
    \item 
    \emph{Orthogonal FBs} result in compact (strictly finite support) FB projectors, very similar to their CLSs. 
    \item
    \emph{Linearly independent FBs} are gapped and characterized by exponentially decaying projectors dressed by algebraic prefactors, due to the nonzero overlap of neighboring CLSs.
    The exponent is related to a localization length scale \(\xi\) and the algebraic prefactor is given by \(r^{-(d-1)/2}\).
    We obtain analytical estimates for \(\xi\) for small band gaps (Fig.~\ref{fig:projector_result_summary} and text below).
    \item 
    \textit{Singular FBs} (\(d\geq 2\)): Linear dependence of the CLS set enforces \(n\)-fold degeneracy of the FB Hamiltonian at a singular location in \(k\) space, typically with \(n=2,3\).
    The flat band touches one (or more) dispersive bands.
    The projector decay is entirely defined by the Taylor expansion of the FB Bloch eigenstates around the degeneracy point.
    The algebraic decay is \(r^{-d}\), unless anisotropic corrections apply (Fig.~\ref{fig:anisotropic_result} and text below).
\end{itemize}
We demonstrate a smooth tuning of an orthogonal to linearly independent to singular flat band and its associated localization length.
We further reveal the subtle emergence of the crossover from the \(1/r^{(d-1)/2}\) decay component of a linearly independent FB projector to the \(1/r^d\) projector decay of a singular FB at a distance of the order of \(\xi\).
We obtain analytical estimates for $\xi$ in this regime, including its anisotropic directional dependence (Fig.~\ref{fig:projector_result_summary}).
We then demonstrate the anisotropic algebraic decay of singular FB models in dimensions \(d=2,3\) using several examples (Fig.~\ref{fig:anisotropic_result}).

\section{Basic setup}
\label{sec:basic_setup}

We consider a finite-range hopping tight-binding Hamiltonian \(H\) on a \(d\)-dimensional lattice \(\Lambda\) with a number \(u\) of sublattices.
In the momentum basis \(\ket{\mu, \bk}\), \(H\) is block-diagonalized into a \(u \times u\) Hamiltonian matrix for each \(\bk\), \(\mathcal{H}(\mu,\nu;\bk) = \bra{\mu, \bk}H\ket{\nu, \bk}\) with \(\mu, \nu\) the row and column orbital indices. 
Diagonalizing \(\mathcal{H}(\bk)\) yields the band structure.
We assume a single perfectly flat band of energy \(E_{\fb}\) with Bloch eigenvector \(\psi_{\fb}(\mu,\bk)\) in the band structure.
For any FB with normalized Bloch eigenvector \(\psi_{\fb}(\mu,\bk)\) there exists a scalar gauge factor \(\alpha(\bk)\) such that the Bloch-CLS (BCLS) \(\phi_{\bcls}(\mu,\bk)\equiv \alpha(\bk)\,\psi_{\fb}(\mu,\bk)\) is a finite Laurent polynomial in \(z_j=e^{ik_j}\)~\cite{read2017compactly}.
We denote by $\ket{\cls^{\br}}$ the CLS centered at $\br$, with the superscript indicating its location.
The CLS at the origin has amplitudes $\phi_{\cls}(\mu,\br)\equiv \braket*{\mu,\br}{\cls^{\mathbf 0}}$, and is given by the inverse lattice Fourier transform of $\phi_{\bcls}(\mu,\bk)$:

\begin{align}
    \label{eq:bcls_fourier}
    \phi_\bcls(\mu,\bk)=\sum_{\br}\phi_\cls(\mu,\br)\,e^{-i\bk\cdot\br}.
\end{align}
We repeat that \(\phi_{\cls}(\mu, \br)\) is compact and has finite support.
The gauge \(\alpha(\bk)\) is chosen such that the components of BCLS share no nontrivial common Laurent polynomial factor other than 1, or an overall phase.
We define such a BCLS as irreducible, corresponding in real space to an irreducible CLS~\cite{maimaiti2017compact}.
Irreducibility is important in flat band classification, since any reducible nonorthogonal CLS can be generated from an irreducible one, and only the algebraic properties of irreducible CLS affects the properties of the projector.

With a proper gauge choice, we can assume \(\alpha(\bk) \geq 0\) and from Eq.~\eqref{eq:bcls_fourier} we obtain
\begin{gather}
    \label{eq:alpha2_2}
    \alpha^{2}(\bk) =\sum_{\br\in\Lambda} \braket*{\cls^{\mathbf 0}}{\cls^{\br}}\,e^{i\bk\cdot\br} = \phi^\dagger_\bcls(\bk) \phi_\bcls(\bk).
\end{gather}
It is clear that \(\alpha^{2}(\bk)\) is also a finite Laurent polynomial.
\(\alpha(\bk)\) is a signature for the threefold classes introduced above:
(i) \emph{orthogonal} FBs: \(\alpha(\bk)\) is constant;
(ii) \emph{linearly independent} FBs: \(\alpha(\bk)\) varies with \(\bk\) but is gapped away from zero for all \(\bk\);
(iii) \emph{singular} FBs: \(\alpha(\bk)\) varies with \(\bk\) with no gap from zero such that \(\alpha(\bk_{0})=0\) for some \(\bk_0\).

\section{Real space behavior of flat band projectors}
\label{sec:real_space_fb}

The real space and momentum space representation of projection onto the FB are given by:
\begin{gather}
    \label{eq:projector_fourier}
    P(\mu, \nu;\br) = \frac{1}{(2\pi)^d}\int_{\bz} \mathcal{P}(\mu,\nu;\bk)e^{i\bk \cdot \br}d^d\bk \\    \label{eq:projector_simplified_definition}
    \mathcal{P}(\mu,\nu;\bk) = \frac{\phi^*_\bcls(\mu, \bk) \phi_\bcls(\nu,\bk)}{\alpha^2(\bk)}
\end{gather}
where \(\bz\) refers to Brillouin zone.
Such projectors are localized in real space, since assuming \(|\mathcal{P}(\mu,\nu;\bk)| < \infty\) results in \(P(\mu,\nu;\br) \to 0\) when \(\br \to \infty\) due to Riemann-Lebesgue lemma~\cite{bochner1949fourier}.
The asymptotic decay of \(P(\mu,\nu; \br)\) for large \(\br\) depends on the FB class.
For orthogonal flat bands the integrand in Eq.~\eqref{eq:projector_fourier} is given as a finite Fourier series of the harmonics in Brillouin zone, since \(\alpha(\bk) = 1\). 
The projector in real space is therefore compact (Appendix.~\ref{app:cls_and_bcls}, and earlier results for \(d=1\) in Ref.~\onlinecite{sathe2021compactly}).

In the linearly independent case the long-distance behavior of the integral in Eq.~\eqref{eq:projector_fourier} can be understood by interpreting \(\mathcal{P}(\mu,\nu;\bk)\) in Eq.~\eqref{eq:projector_simplified_definition} as a Green function at \(E=0\). 
The asymptotics of the projector is therefore similar to that of the Green's function evaluated outside the band.
To apply the saddle point analysis to Eq.~\eqref{eq:projector_fourier}, we use the Schwinger parametrization
\begin{align}
    \frac{1}{\alpha^2(\bk)}=\int_0^{\infty} e^{-s\alpha^2(\bk)}\,ds.
\end{align}
After the rescaling \(s=rt\), we obtain
\begin{gather}
    \label{eq:projector_real_saddle_expression}
    P(\mu,\nu;\br)=r\int_0^{\infty}dt\int d^d\bk\,
    N(\bk)\,e^{r f(t,\bk)},\\
    \notag
    f(t,\bk)=-t\alpha^2(\bk)+i\bk\cdot\hat\br,
\end{gather}
where \(N(\bk)\) is the numerator of Eq.~\eqref{eq:projector_fourier} and \(\br=r\hat\br\).
This is the standard form for the saddle point analysis.

For a linearly independent flat band, the relevant saddle is reached only after deforming the momentum contour into complex momentum space, so that it passes through a zero of \(\alpha(\bk)\)~\cite{cloizeaux1964energy}.  
The saddle with the smallest \(|\Im \bk|\) gives the slowest decay.  
Expanding around this saddle and evaluating the resulting Gaussian integral (Appendix~\ref{app:OZ_decay}), we obtain the generic Ornstein-Zernike form
\begin{gather}
    P(\mu,\nu;\br)\sim r^{-\frac{d-1}{2}}\exp\bigg(-\frac{r}{\xi_{\hat\br}}\bigg),
\end{gather}
with a direction-dependent localization length \(\xi_{\hat\br}\).

In the 1D case, the above saddle point argument is unnecessary, and one can evaluate     Eq.~\eqref{eq:projector_fourier} exactly using contour integration, and the localization length \(\xi\) is an explicit function of the overlap \(\sigma\), for CLSs satisfying \(\braket{\cls_j}{\cls_i} = \delta_{i,j} + \sigma \delta_{i,\pm 1}\), which was derived in Refs.~\onlinecite{chasemayoral2024compact, marques2024impurity}. 

It is instructive to consider two limiting cases of linearly independent flat bands when tuned into singular and orthogonal FBs.
We remind that such a tuning can be easily obtained by smoothly varying the amplitudes of the CLS accordingly.

For the case of a linearly independent FB tuned into an orthogonal one, the overlap between CLS tends to zero, with localization length \(\xi\) consequently also going to zero. 
The bulk part of the projector is dominated by the compact projector in the orthogonal limit, as discussed in  Appendix~\ref{app:orthogonal_fb}.

\subsection{Nearly singular limit}
\label{sec:real_space_fb:nearly_singular}

For a linearly independent FB close to a singular one, the minimum
\(\Delta\equiv \min_{\bk}\alpha^2(\bk)\) becomes small.
Let \(\bk_1\) denote the momentum where this minimum is attained.
Note that \(\Delta\) is not necessarily the actual band gap of the full Hamiltonian.
In the nearly singular regime, the relevant contribution to the projector is controlled by momenta near \(\bk_1\), and both denominator and numerator can be expanded around this point.

In the generic case,
\begin{align}
    \label{eq:alphasq_quadratic}
    \alpha^2(\bk)\approx \Delta+(\bk-\bk_1)^T B(\bk-\bk_1),
\end{align}
with \(B\) the positive definite curvature matrix for \(\alpha^2(\bk)\).
The same local form also determines the relevant complex saddle, which lies close to \(\bk_1\), and yields an analytic expression [Eq.~\eqref{eq:saddle_analytical_solution} of Appendix] for the direction-dependent localization length:
\begin{align}
    \label{eq:xi_analytical}
    \xi^{-1}_{\hat\br} = \sqrt{\Delta} \sqrt{\hat \br^T B^{-1} \hat\br}.
\end{align}

More importantly, the projector integral itself can be approximated by its infrared contribution around \(\bk_1\).
The infrared form captures not only the saddle point asymptotics, but also the intermediate crossover regime.
As a result, the projector displays singular-FB-like decay \(P(\br)\sim r^{-d}\) for \(r\ll \xi_{\hat\br}\), before crossing over to the asymptotic Ornstein--Zernike form for \(r\gg \xi_{\hat\br}\).

Thus, the generic singular flat band is recovered as the limit \(\xi\to\infty\), in which the crossover scale is pushed to infinity and the algebraic decay remains as the asymptotic behavior.
The localization length scales as \(\xi_{\hat\br}\propto \Delta^{-\gamma}\), with \(\gamma=1/2\) for quadratic touchings and \(\gamma=1\) for linear touchings.

\section{Numerical simulations}
\label{sec:numerical_simulations}

We now confirm numerically the validity of the above statements.
As an example we consider a (\(d = 2\)) FB model with 2 sublattices, (\(u = 2\)), based on a modified checkerboard lattice and with a CLS parametrized by parameter \(A\), \(0 \leq A \leq 1\), encompassing all the three classes:
orthogonal ((\(A = 0\)), linearly independent (\(0<A<1\)), singular (\(A = 1\)) (Fig.~\ref{fig:projector_result_summary}(b)). 
The CLS reads as follows:
\begin{gather}
    \label{eq:square_cls} 
    \ket{\cls^\br} = A(\ket{1,\br} - \ket{2,\br}) + \ket{1, {\br+\hat{\mathbf{x}}}} - \ket{2, {\br - \hat{\mathbf{y}}}}
\end{gather}
where we have omitted the normalization factor \(\frac{1}{\sqrt{2A^2 + 2}}\).
From this expression it is straightforward to obtain the BCLS \(\phi_\bcls(\bk)\):
\begin{gather}
    \label{eq:square_bcls} 
    \phi_\bcls(\bk)=
    \begin{bmatrix}
        A + e^{ik_x}\\[2pt]
        -\,(A+e^{-ik_y})
    \end{bmatrix},
\end{gather}
and the projector~\eqref{eq:projector_fourier} in real space.
For \(A = 0\) the CLS set is orthogonal, while for \(A = 1\) the flat band is singular, with the touching at \(k_x = k_y = \pi\).
The above CLS alone is enough to construct the projector and analyze its properties.
One of the many possible Hamiltonians is given by the spectral decomposition~\cite{graf2021designing, hwang2021general}: \(\mathcal H(\bk) = \alpha^2(\bk)I - \phi_{\bcls}(\bk)\phi^\dagger_{\bcls}(\bk)\).
Its lattice structure and the band structure are shown in Fig.~\ref{fig:projector_result_summary}(a)-(b).
The two bands are given as \(E_{\fb} = 0\) and \(E_{\db} = \alpha^2(\bk)\).

Next we compute numerically the real space projector in \(x\)-direction \(P(1, 1; x)\) obtained from the CLS~\eqref{eq:square_cls} using Eq.~\eqref{eq:projector_fourier}.
To verify the asymptotic decay of the projector, we assume the OZ-like ansatz: \(P(x) \sim x^{a} e^{-x/\xi_{\hat{x}}}\), where \(\xi_{\hat{x}}\) is the localization length in the x-direction, and check its validity [Fig.~\ref{fig:projector_result_summary}(e-f)].
To extract parameters \(\xi_{\hat x}\) and \(a\), we devised a ratio-test type analysis~\cite{arfken2011chapter1} using the ratio 
\begin{align}
    \label{eq:ratio_test}
    f(x) = \frac{P(1, 1;x+1)}{P(1,1;x)},
\end{align} 
[Fig.~\ref{fig:projector_result_summary}(c)].
With the OZ ansatz for the projector, the asymptotic behavior of \(f(x)\) for large \(x\) is expected to be: 
\begin{align}
    f(x) \sim \exp(-\xi^{-1}_{\hat{x}})(1 + ax^{-1}).
\end{align}
Since the algebraic part is linear in \(x^{-1}\), we used a linear fit with respect to \(x^{-1}\) for large \(x\).
The slope and the \(y\)-intercept of the fit is then used to extract the parameters \(a\) and \(\xi_{\hat x}\).

The OZ ansatz is confirmed by the linear behavior of \(f(x)\) for large \(x\) (small \(x^{-1}\)) as shown in the inset of panel (e) of Fig.~\ref{fig:projector_result_summary}.
The extracted length \(\xi_{\hat x}\) is shown in panel (f), and agrees well with theory (Sec.~\ref{sec:real_space_fb:nearly_singular} and Appendix~\ref{app:OZ_decay}).
The panels (c) and (d) show the behavior of the algebraic prefactor, \(P(x)e^{-x/\xi_{\hat x}}\).
For large \(x\), we observe the expected OZ decay, \(x^{-1/2}\). 
Upon approaching the singular limit \(A = 0.9\), panel (d), we observe a crossover around \(x\approx \xi\) from an emergent algebraic decay of the singular case, \(x^{-2}\) for \(x < \xi\), to the OZ decay, \(x^{-1/2}\) for \(x > \xi\), confirming our analytical predictions of Sec.~\ref{sec:real_space_fb:nearly_singular}.

\section{Anisotropic \texorpdfstring{\(d = 2\)}{d=2} singular flat bands}
\label{sec:anisotropic_singular_fb}

Anisotropy of the hoppings can result in singular FBs with an anisotropic algebraic decay.
Indeed, for a singular FB the directional anisotropy of the BCLS \(\phi_\bcls(\nu,\bk)\) near the touching point \(\bk_0\) alters the projector’s asymptotic decay.
To illustrate this, we consider the Taylor expansion of the BCLS in \(\bq = \bk - \bk_0\) around the touching point \(\bk_0\): 
\begin{align}
    \phi_{\mathrm{BCLS}}(\bk_0+\bq)= i\,b\,\bq+O(|\bq|^2),
\end{align}
where \(b\) is a \(u\times d\) Jacobian matrix (complex in general, for complex CLS).
We consider the real, \(2u\times d\), matrix
\begin{align}
    \tilde b =
    \begin{pmatrix}
        \Re b\\
        \Im b
    \end{pmatrix}.
\end{align}
Then we have
\begin{align}
    \notag \alpha^2(\bk_0+\bq)
    &= \phi^\dagger_{\mathrm{BCLS}}(\bk_0+\bq)\phi_{\mathrm{BCLS}}(\bk_0+\bq)\\
    &= \bq^T B \,\bq + O(|\bq|^3),\\
    B &= \tilde b^{T}\tilde b = (\Re b)^T(\Re b)+(\Im b)^T(\Im b).
\end{align}
A generic singular flat band is characterized by a non-singular matrix \(B\) when expanded around its touching point. The quadratic form \(\bq^T B \,\bq\) controls the touching for all directions producing the default algebraic decay \(r^{-d}\).
By contrast, a singular matrix \(B\) leads to an anisotropic algebraic decay of the flatband projector.
For the null directions of a real valued \(\bq_{\parallel}\): \(B\bq_{\parallel}=0\), higher-order terms in \(\bq\) control the decay of the projector.
For a real-valued CLS, \(b\) is real and replaces \(\tilde b\) in the above derivation.

\begin{figure}[htpb]
    \centering
    \includegraphics[width=0.99\linewidth]{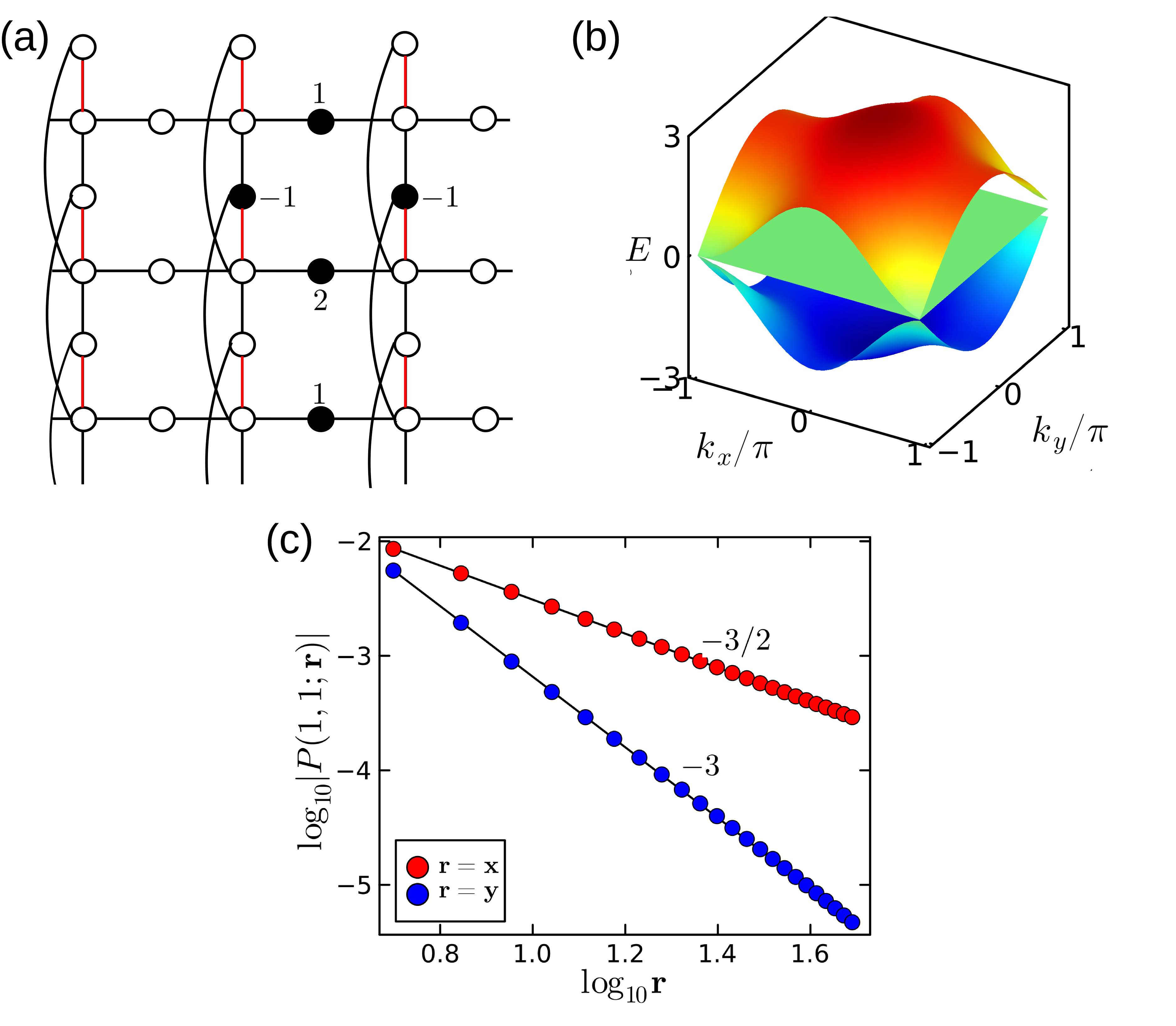}
    \caption{
        A generalized anisotropic Lieb lattice which hosts an anisotropic singular flat band.
        (a) Lattice, the black lines are hoppings with strength 1, and the red lines are hoppings with strength 2.
        The CLS corresponds to the black colored circles [Eq.~\eqref{eq:nongeneric_bcls}].
        (b) Band structure of the lattice (a). 
        (c) Decay of the projector \(|P(1, 1; \br)|\) vs. \(\br\) in log-log scale.
        Red: \(\mathbf{x}\) direction, linear fitting shows exponent \(-3/2\), Blue: \(\mathbf{y}\) direction, linear fitting shows exponent \(-3\).
    }
    \label{fig:anisotropic_result}
\end{figure}

We illustrate the anisotropic decay with the following example of a modified 2D Lieb lattice Fig.~\ref{fig:anisotropic_result}(a). 
The anisotropy is due to additional longer range hoppings in the \(y\)-direction, and the anisotropic CLS is shown with filled circles.
Since the chiral symmetry is not violated, the chiral flat band resides at energy \(E=0\)~\cite{ramachandran2017chiral} and experiences a linear band touching with two dispersive bands [Fig.~\ref{fig:anisotropic_result}(b)].
Note that we only need the BCLS to analyze the properties of a projector, since the full Hamiltonians can be constructed using the methods of Refs.~\onlinecite{maimaiti2021flatband,graf2021designing,hwang2021general}.
We consider first a 2D singular FB, where the BCLS is given as 
\begin{align}
    \label{eq:nongeneric_bcls}
    \phi_{\mathrm{BCLS}} =
    \begin{bmatrix}
        -e^{ik_x}-1\\
        e^{ik_y} + e^{-ik_y} +2
    \end{bmatrix}
\end{align}
Near the touching point, \(\bk =(\pi, \pi)\) and \(q_i = k_i + \pi\), \(B=\mathrm{diag}(1,0)\) and we have \(P(1,1;\bk) \sim q_x^2/(q_x^2+q_y^4)\) that is invariant under rescaling \((q_x,q_y)\to(\lambda q_x,\lambda^{1/2}q_y)\).
Consequently, in real space, \(P(\mu, \nu;\br)\sim |\br|^{-3/2}\), except for the \(y\)-direction, where \(P(\mu, \nu;\mathbf{y}) \sim |\mathbf{y}|^{-3}\).
This is verified numerically as shown in Fig.~\ref{fig:anisotropic_result}(c).
We construct the Lieb-type lattice Hamiltonian \(\mathcal H(\bk)\) that hosts the BCLS in Eq.~\eqref{eq:nongeneric_bcls},
following the construction method of Ref.~\onlinecite{graf2021designing}; see Fig.~\ref{fig:anisotropic_result}(a).
The corresponding band structure is shown in Fig.~\ref{fig:anisotropic_result}(b).
The real-space projector is obtained via a straightforward Fourier transform of the projector in the momentum space, and the result appears in Fig.~\ref{fig:anisotropic_result}(c).
All calculations use system size \(L = 3000\).
A power law fit (linear fit in log-log) indicates that along the \(\mathbf{x}\) direction the projector decays as \(|\mathbf{x}|^{-3/2}\) (red), whereas along the \(\mathbf{y}\) direction it decays as \(|\mathbf{y}|^{-3}\) (blue).

It follows that for \(d > 2\), and real valued CLS with a number of sublattices \(u < d\), the above band touching anisotropy is enforced:
for real-valued CLS, \(\tilde b\) is replaced by \(b\), where \(b\) is a real \(u\times d\) matrix.
Since \(\mathrm{rank}(b)\le u\), \(\mathrm{rank}(B) \le u < d = \dim(B)\) and \(B\) is necessarily singular.

Indeed, consider a BCLS with two sublattices (\(u = 2\) in \(d = 3\)): 
\begin{align}
    \phi_{\mathrm{BCLS}}=
    \begin{bmatrix}
        e^{ik_x}+e^{ik_y}-2\\
        e^{ik_z}-1
    \end{bmatrix}
\end{align}
At the touching point \(\bk = 0\), the matrix \(B = b^\dagger b\) has to be singular since it is rank deficient: \(b\) is a \(2\times 3\) matrix.
Here the real space decay is \(|\br|^{-5/2}\) along generic directions, and \(|\br|^{-5}\) along the nullspace directions of \(B\).
If we allow for complex valued CLS, we  can evade the singularity of \(B\) even when \(u<d\). 
For \(\phi_{\mathrm{BCLS}}=[(e^{ik_x}-1)+i(e^{ik_y}-1),e^{ik_z}-1]^T\) the correct rank of \(b\) is three for real variables, restoring homogeneous decay \(r^{-d}\) along all directions (Appendix.~\ref{app:nongeneric_sfb}).

The above results on the decay of FB projectors apply to the slowest asymptotic decay only, allowing for special directions with faster decay.
A simple example is the Lieb lattice: the chiral symmetry enforces zero BCLS amplitude on minor sublattices, so if either \(\mu\) or \(\nu\) is chosen as those, the projector is identically zero.
Directional dependence can be modified by symmetry: it might force the denominator of \(\mathcal{P}(\mu,\nu;\bk)\) to cancel out with the numerator for some direction(s) \(\bk\), making the projector compact along that specific direction.

\section{Conclusion}

We showed how the algebra of compact localized states (CLS) is directly related to the qualitatively different long distance behavior of the flat band (FB) projector in real space.
Orthogonal flat bands yield strictly compact projectors, while linearly independent flat bands exhibit Ornstein-Zernike decay with a direction-dependent localization length.
Singular flat bands display algebraic decay \(r^{-d}\) in general, but strongly anisotropic power law decay with different powers is possible for anisotropic cases.
We have demonstrated a crossover: as a linearly independent FB approaches a singular one, the projector crosses over from emergent algebraic decay at short distances to the asymptotic Ornstein-Zernike form at long distances.
In particular, the localization length \(\xi \gg 1\) sets the crossover scale: for distances \(r\ll \xi\), the projector follows the same algebraic decay as in the singular flat band limit, so that on such length scales the system can be treated effectively as a singular flat band, while only for \(r\gg \xi\) does the true asymptotic OZ behavior emerge.
We also identify an anisotropic singular FB class in which anisotropy of the band touching combined with higher order touching controls the decay:
for generic spatial directions the projector follows a slower algebraic decay, than the generic singular FBs with \(r^{-d}\), while along special axes it shows a faster algebraic decay than the generic case.
We believe that this new non-generic class can be relevant to perturbed FBs, e.g. superconductivity, optical and transport response, and disordered systems.
Local excitations of such lattices which resonate with the FB energy will generate a spatial response which follows the decay of the above real space projectors.

For a nearly singular limit, Sec.~\ref{sec:real_space_fb:nearly_singular}, we only considered the case of generic singular flat bands, but our analysis could be straightforwardly extended to the case of proximity to anisotropic singular flatbands discussed in Sec.~\ref{sec:anisotropic_singular_fb}.

It is also instructive to discuss the limitations of our results.
Our classification scheme relies on the existence of compact localized states (CLS) ensuring the flatband projector can be given by Eq.~\eqref{eq:projector_simplified_definition}.
The existence of CLS is guaranteed for Hamiltonians with strictly finite-range/compact hopping and a finite number of bands (that is, finite number of sites in a unit cell).
For infinite-range/non-compact hopping, the existence of CLS is not guaranteed in general, and therefore our classification may not be applicable.
One such example is provided by the Wannier-Stark or Floquet flatbands~\cite{mallick2021wannier} which have finite range hopping and flat bands, but an infinite number of bands.

We acknowledge a recent development~\cite{lee2025embedding}, where the authors independently study a length scale associated with a FB (coinciding with the length scale defined in this manuscript for the linearly independent case) and show its relevance to superconductivity.
This interesting work further supports the importance of flatband projector length scales.

\begin{acknowledgments}
    The authors acknowledge financial support from the Institute for Basic Science (IBS) in the Republic of Korea through Project No. IBS-R024-D1.
\end{acknowledgments}

\appendix

\section{CLS classification}
\label{app:cls_classification}

\subsection{Compact localized states (CLS) and the Bloch-CLS}
\label{app:cls_and_bcls}

Let us consider a flat band (FB) tight-binding Hamiltonian \(H\) defined on a lattice \(\Lambda\), with \(u\) sublattices (\(u\)-orbitals), and its \(u\times u\) momentum-space block \(\mathcal H(\bk)\) discussed in the main text.
The BCLS is defined as the non-normalized FB eigenstates with a proper gauge, \(\phi_\bcls(\mu,\bk)=\alpha(\bk)\psi_\fb(\mu,\bk)\), such that we can write \(\phi_\bcls(\mu,\bk)\) as a finite Fourier series:
\begin{align}
    \phi_\bcls(\mu,\bk)=\sum_{\br}\phi_\cls(\mu,\br)\,e^{-i\bk\cdot\br},
\end{align}
where \(\phi_\cls(\mu,\br)=0\) for \(|\br|>r_0\).
Thus, for the CLS located at the unit cell \(\br_0\), the amplitude is given by
\begin{align}
    \phi_\cls(\mu,\br;\br_0) = & \,\,\phi_\cls(\mu,\br-\br_0) \notag \\
    =& \int \phi_\bcls(\mu,\bk)\,e^{i\bk\cdot(\br-\br_0)}\,d^d\bk.
\end{align}
The CLS is an eigenstate because it is a linear combination of BCLS.

\subsection{Irreducible CLS}
\label{app:irreducible_cls}

To obtain the irreducible CLS, one ensures a BCLS \(\phi_\bcls(\bk)\) does not contain a common finite Fourier series factor other than 1.
To understand the motivation of this definition, let us suppose we have a suboptimal (reducible) BCLS as \(\phi'_\bcls(\bk)=(1+e^{ik_1})\phi_\bcls(\bk)\), where the \(\phi_\bcls(\mu,\bk)\) on the right-hand side is the irreducible BCLS.
For this suboptimal (reducible) choice, we obtain
\begin{align}
    \phi'_\cls(\mu,\br;\br_0)
    &= \int (1+e^{ik_1})\,e^{i\bk\cdot(\br-\br_0)}\,\phi_\bcls(\mu,\bk)\,d^d\bk \notag \\
    &= \phi_\cls(\mu,\br;\br_0)+\phi_\cls(\mu,\br;\br_0-\hat{\mathbf{x}}),
\end{align}
resulting in a linear combination of the irreducible CLSs located at \(\br_0\) and \(\br_0-\hat{\mathbf{x}}\).

From this irreducibility condition one can map the projector properties and algebraic properties of the CLS one-to-one.

\subsection{Overlap between CLS and the norm square of BCLS}
\label{app:overlap_cls}

The norm square of BCLS, \(\alpha^2(\bk)=\phi_\bcls^\dagger(\bk)\phi_\bcls(\bk)\), is a finite Fourier series, and the Fourier coefficients are the overlaps between CLS.
We have
\begin{align}
    \notag \alpha^2(\bk)
    &= \sum_{\br,\br'} \phi_\cls^*(\mu,\br)\,\phi_\cls(\mu,\br')\,e^{i\bk\cdot(\br-\br')}\\
    \notag &= \sum_{\Delta}\bigg[\sum_{\br}\phi_\cls^*(\mu,\br)\,\phi_\cls(\mu,\br-\Delta)\bigg]e^{i\bk\cdot\Delta}\\
    \label{eq:alphasq_overlap} &= \sum_{\Delta}\braket{\cls_0}{\cls_{\Delta}}\,e^{i\bk\cdot\Delta}.
\end{align}
It follows that if \(\alpha^2(\bk)=\mathrm{const.}\), the CLS are orthogonal.

\section{Orthogonal flat bands}
\label{app:orthogonal_fb}

For an orthogonal flat band we have \(\alpha^2(\bk)=1\), so the projector in real space is
\begin{gather}
    \label{eq:projector_fourier_ortho_int}
    P(\mu,\nu;\br)=\frac{1}{(2\pi)^d}\int_{\text{BZ}} \phi^*_\bcls(\mu,\bk)\,\phi_\bcls(\nu,\bk)\,e^{i\bk\cdot\br}\,d^d\bk.
\end{gather}
The Fourier transform results in
\begin{gather}
    \label{eq:projector_fourier_ortho}
    P(\mu,\nu;\br)=\sum_{\br'}\phi^*_\cls(\mu,\br')\,\phi_\cls(\nu,\br'-\br),
\end{gather}
which is compact.

\section{Non-orthogonal flat bands}
\label{app:OZ_decay}

We provide here the details of the derivation of the Ornstein--Zernike decay for the projectors of exponential flat bands.

The saddle point of Eq.~\eqref{eq:projector_real_saddle_expression} satisfies:
\begin{align}
    \notag \partial_t f(t,\bk)&=-\alpha^2(\bk)=0,\\
    \label{eq:saddle_point}
    \nabla_\bk f(t,\bk)&=-t\nabla_\bk\alpha^2(\bk)+i\hat{\br}=0.
\end{align}

Let us start from Eq.~\eqref{eq:saddle_point},
First, we note that \(|\alpha(\bk)|=0\) cannot occur for real \(\bk\) for nonsingular flat bands, and we must shift the integration contour so that it crosses the zero.
Let us assume that we have solved Eq.~\eqref{eq:saddle_point} and have obtained the saddle point(s) \((t_0(\hat\br),\bk_0(\hat\br))\) (in case of multiple solutions, the relevant one has the smallest \(|\Im \bk_0|\)).
Let us write \(\bk_0(\hat\br)=\bk_0\) and \(t_0(\hat\br)=t_0\).
After locating the saddle, expand the exponent of Eq.~\eqref{eq:projector_fourier} to quadratic order in \((t,\bk)\). Let \(t'=t-t_0\).
We have
\begin{align}
    f(t,\bk)=i t'(\bq\cdot \vec{v})+\bq^T B \bq+O(|\bq|^3),
\end{align}
where \(\bq=\bk-\bk_0\), \(\vec{v}=\hat \br/t_0\), and \(B_{ij}(\bk_0)=\tfrac{t_0}{2}\,\partial_{k_i}\partial_{k_j}\alpha^2(\bk)\big|_{\bk=\bk_0}\).
Importantly, we assume \(B=B(\bk_0)\) to be nonsingular at the saddle.
We also consider the case \(N(\bk)\approx N_0\).
Completing the square, we have
\begin{align}
    f(t,\bk)\approx \bq'^T B \bq' - t'^2\,\vec{v}^{T} B^{-1}\vec{v}.
\end{align}

The integrand is now a product of two independent Gaussians.
The \(\bq'\) integration gives \((2\pi)^{d/2}[\det B]^{-1/2}(rt_0)^{-d/2}\).
The \(t'\) integration contributes \((2\pi r^{-1}t_0^{\,2}/c_{\hat{\br}})^{1/2}\), where \(c_{\hat{\br}}=\hat{\br}^{T}B^{-1}\hat{\br}\).
Multiplying the two factors gives \(r^{-d/2-1/2}\) decay.
Combining with the outside \(r\) prefactor, we obtain
\begin{align}
    P(\mu,\nu;\br) \sim \frac{e^{-r/\xi_{\hat{\br}}}}{r^{(d-1)/2}},
\end{align}
which is the OZ decay~\cite{ornstein1914accidental,michta2021asymptotic}.

We have assumed a non-vanishing zeroth order of \(N(\bk)\) near the saddle \(\bk_0(\hat\br)\) in the derivation.
If this is not the case, one can apply Feynman's trick: if \(N(\bk)\) starts from first order near the saddle, the \(\bk\)-prefactor becomes derivatives in real space.
Therefore, the algebraic prefactor becomes \(r^{-(d+1)/2}\).
Also, symmetry often forces \(N\big(\bk_0(\hat\br)\big)=0\), i.e., the integral vanishes along certain lines.
In this case, the projector is compact in that particular \(\hat \br\) direction.

A more subtle situation arises when \(B\) is singular.
In that case, the minimum of \(\alpha^2(\bk)\) is strongly anisotropic, with some directions not controlled by the quadratic form.
The standard Gaussian saddle point expansion around the minimum is then no longer sufficient, and the asymptotic decay must be analyzed separately for each type of anisotropy.

\section{Singular flat bands}
\label{app:power_law}

\subsection{Generic case}

 Here we consider the limit of the nonsingular flat band becoming a singular one, i.e., the band gap between a flat and dispersive band vanishes, \(\Delta\to 0\).
For any \(\Delta>0\) the \(r\to\infty\) results of Section~\ref{app:OZ_decay} still apply for
\begin{align}
    \label{eq:Pr}
    P(\mu,\nu;\br)=\int \frac{N(\bk)\,e^{i\bk\cdot\br}}{\alpha^2(\bk)}\,d^d\bk.
\end{align}
As \(\Delta\to 0\), the relevant zero of \(\alpha^2(\bk)\) approaches the real axis, and simultaneously so does the zero of \(N(\bk)\).
Exactly at \(\Delta=0\), the situation depends on the dimension of space.
In 1D, this leads to the cancellation of the zeros, producing an orthogonal flat band.
However, in \(d>1\), while the zeros also cancel out, there might be a residual direction dependence in approaching the zero, which is not possible in 1D.
For small enough, but nonzero \(\Delta\), this leads to the emergence of the singular flat band projector decay, \(r^{-d}\), followed by the Ornstein--Zernike decay.
This intermediate asymptotics is not captured by the saddle-point derivation of Section~\ref{app:OZ_decay}.
However, for small \(\Delta\), we can directly evaluate the saddle-point integral~\eqref{eq:Pr} by using an approximation of the denominator \(\alpha^2(\bk)\).
As explained in the main text, in the nearly singular case, \(\alpha^2(\bk)\) is approximated as Eq.~\eqref{eq:alphasq_quadratic} near its minimum \(\bk_1\).
Here, \(B\) is a positive semidefinite matrix since it is expanded at the minimum, but for now we will assume that \(B\) is nonsingular (positive definite).
To find the saddle, invoke Eq.~\eqref{eq:alphasq_quadratic} directly into Eq.~\eqref{eq:saddle_point}, to obtain
\begin{align}
    \notag \bk_0 &= \bk_1 + i\frac{B^{-1}\hat\br}{2t_0}\\
    \label{eq:saddle_analytical_solution}
    t_0 &= \frac{\sqrt{\hat \br^T B^{-1}\hat\br}}{2\sqrt\Delta}.
\end{align}
This gives us the analytical form of the localization length in Eq.~\eqref{eq:xi_analytical}.

However, as explained in the Main, in the nearly singular FB, one can directly evaluate the integral, to obtain the intermediate real space decay beyond saddle point analysis.
We show the derivation below.

For small \(\Delta\) one can expand the numerator \(N(\bk)\) as
\begin{align}
    N(\bk)\approx a_0 + \vec a \cdot \bq + \bq^T A\bq.
\end{align}
where \(a_0 = N(\bk_1)\), \(\vec a_1 = \nabla N(\bk)|_{k_1}\), and \(A_{ij} = \partial_{k_i}\partial_{k_j}N|_{\bk_1}/2\).
To obtain the intermediate decay, we integrate \(\mathcal P(\mu,\nu;\bk)\) without the Schwinger parameterization:
\begin{align}
    \label{eq:pkq3}
    \mathcal P(\mu,\nu;\bk) & =\frac{a_0+\vec{a}\cdot\bq+\bq^T A\bq}{\Delta+\bq^T B\bq} + O(q^3) = \\
    & = P_{\mathrm{IR}}(\mu,\nu;\bk)+O(q^3). \notag
\end{align}
The band gap \(\Delta\) and matrix \(B\) control the anisotropic length scale \(\xi\)~\cite{rhim2021singular,ashcroft1976solid_appendixE}.
The decay of the real-space projector is given by the Fourier transform of the above expression and is controlled by its small-\(\bq\) behavior:
\begin{align}
    P(\br)\sim \int_{\mathbb{R}^d} P_{\textrm{IR}}(\bq)\,e^{i\bq\cdot\br}\,d^d\bq.
\end{align}
Since we have assumed a nonsingular \(B\), consider the transformation \(U^T B U=\Lambda=\mathrm{diag}(\{\lambda_i\})\) and define \(\bq'=\Lambda^{1/2}U^T\bq\), so that \(\bq^T B \bq=q'^2\).
We then have
\begin{align}
    P(\mu,\nu;\br)\sim \det(B)^{-1/2}\int \frac{a_0+\vec a'\cdot \bq' + \bq'^T A' \bq'}{q'^2+\xi^{-2}}\,e^{i\bq'\cdot\br'}\,d^d\bq',
\end{align}
where
\begin{gather}
    \notag
    \vec a'=\Lambda^{-1/2}U^{\mathrm T}\vec a,\quad
    A'=\Lambda^{-1/2}U^{\mathrm T} AU\Lambda^{-1/2},\\
    \Delta=\xi^{-2}, \quad 
    \br'=B^{-1/2}\br.
\end{gather}
The expression in the denominator is well known and appears in various settings (screened/Yukawa potential~\cite{ashcroft1976solid_chapter17}, correlation length near \(T_c\)~\cite{goldenfeld2018lectures}); \(\Delta\) is often referred to as the “mass gap”.
The Fourier transform of the denominator alone is given by \(\rho(r'/\xi)=K_{(d-2)/2}(r'/\xi)/r'^{(d-2)/2}\), where \(K_{(d-2)/2}\) is the modified Bessel function of the second kind, while the numerator produces derivatives in real space.
Overall,
\begin{align}
    \label{eq:pr_ns}
    \notag P(\mu,\nu;\br)\sim &\,a_0\,\rho\!\left(\frac{r'}{\xi}\right) + \sum_{i}a'_i\,\partial_{r'_i}\rho\!\left(\frac{r'}{\xi}\right) + \\
    &\sum_{ij} A'_{ij}\,\partial_{r'_i}\partial_{r'_j}\rho\!\left(\frac{r'}{\xi}\right).
\end{align}
This directional dependence of projector decay is confirmed for a two-band square-lattice flat band (see the main text), as shown in Fig.~\ref{fig:endmatfig}(b).

Excluding the asymptotic exponential decay, for \(r'\ll \xi\) we have
\begin{align}
    \rho\!\left(\frac{r'}{\xi}\right)\sim \frac{1}{r'^{\,d-2}}.
\end{align}
For a nearly singular flat band, \(a_0', \vec a'\to 0\), and the second-derivative term in Eq.~\eqref{eq:pr_ns} gives the dominant contribution, yielding the decay \(r^{-d}\).
For \(r'\gg \xi\) the first term gives the OZ decay, since \(a_0\) is finite.

For a quadratic band touching, the divergence of the length upon approaching the singular limit is \(\xi_{\hat\br}\propto \Delta^{-1/2}\), while for a linear band touching with chiral symmetry (e.g., Lieb lattice), the exponent is \(\gamma=-1\).
A simple way to see this is the following:
Squaring the chiral Hamiltonian block diagonalizes it, and one of the blocks includes a flat band with quadratic band touching with a gap \(\Delta'=\Delta^2\), so that \(\xi\propto \Delta'^{-1/2}=\Delta^{-1}\).

\subsubsection{Example: generalized checkerboard lattice}
\label{app:checkerboard}

Starting from Eq.~\eqref{eq:square_cls} and \eqref{eq:square_bcls} of the main text, we first obtain the dual (dispersive-band) vector as
\begin{gather}
    \phi_{\db}(\bk)=
    \begin{bmatrix}
        A + e^{-ik_y}\\[2pt]
        A + e^{ik_x}
    \end{bmatrix}.
\end{gather}
One readily checks orthogonality and equal norms
\begin{align}
    \phi_\bcls^\dagger(\bk)\,\phi_\db(\bk) &= 0,\\
    \|\phi_\bcls(\bk)\|^2 &= \|\phi_\db(\bk)\|^2 \equiv \alpha^2(\bk),
\end{align}
with
\begin{align}
    \alpha^2(\bk) 
    &= |A+e^{ik_x}|^2 + |A+e^{-ik_y}|^2 \nonumber\\
    &= 2\big(A^2+1\big) + 2A\big(\cos k_x + \cos k_y\big).
    \label{eq:alpha2}
\end{align}
The Hamiltonian that hosts the CLS in Eq.~\eqref{eq:square_bcls} can then be constructed as a rank-one projector onto \(\phi_\db\),
\begin{align}
    \mathcal H(\bk)=\phi_\db(\bk)\,\phi_\db^\dagger(\bk)
    = \alpha^2(\bk)\,\mathbb{I} - \phi_\bcls(\bk)\,\phi_\bcls^\dagger(\bk),
\end{align}
i.e., explicitly
\begin{align}
    \mathcal H(\bk)
    = \begin{bmatrix}
        |A+e^{-ik_y}|^2 & \big(A+e^{-ik_y}\big)\big(A+e^{-ik_x}\big)\\[4pt]
        \big(A+e^{ik_y}\big)\big(A+e^{ik_x}\big) & |A+e^{ik_x}|^2
    \end{bmatrix}.
    \label{eq:H-matrix}
\end{align}
By construction,
\begin{align}
    \mathcal H(\bk)\phi_\bcls(\bk) = 0,\quad
    \mathcal H(\bk)\phi_\db(\bk) = \alpha^2(\bk)\phi_\db(\bk),
\end{align}
so the spectrum consists of a perfectly flat band at \(E_\fb(\bk)=0\) spanned by \(\phi_\bcls\), and a generally dispersive band \(E_\db(\bk)=\alpha^2(\bk)\) spanned by \(\phi_\db\).

The band gap is therefore
\begin{align}
    \Delta = \min_{\bk}\,\alpha^2(\bk)
    = 2\big(|A|-1\big)^2,
\end{align}
attained at \(\bk=(\pi,\pi)\) for \(A>0\) and at \(\bk=(0,0)\) for \(A<0\). Hence the gap closes at \(A=\pm1\) and opens quadratically in \(|A|-1\), providing a single-parameter tunable gap.

\begin{figure}[t]
    \centering
    \includegraphics[width=0.5\linewidth]{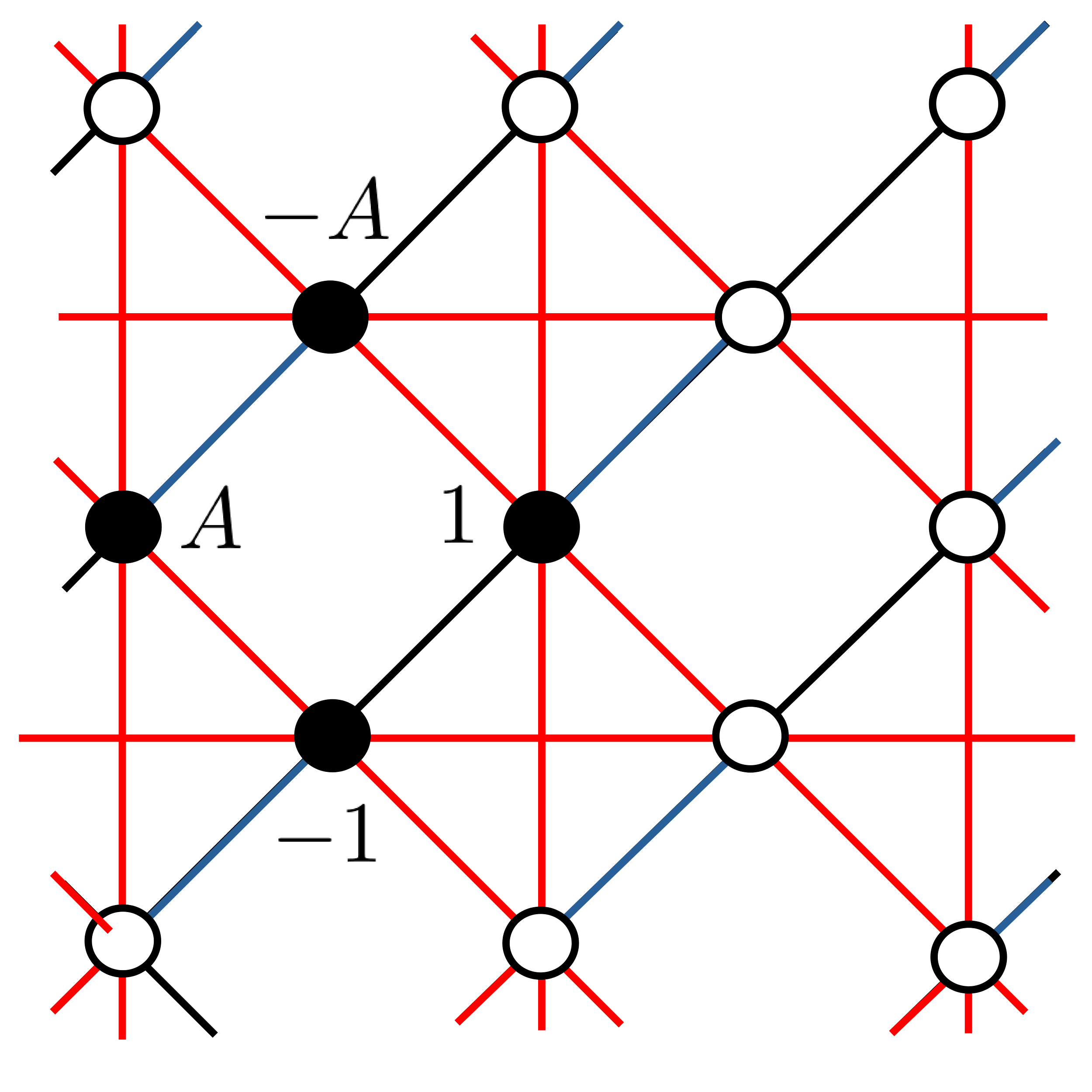}
    \caption{
        Generalized checkerboard lattice. The CLS are black circles.
        Hoppings: black lines \(=1\), red lines \(=A\), blue lines \(=A^2\).
    }
    \label{fig:generalized_checkerboard_lattice}
\end{figure}

We denote by \(t_{\mathbf R,\alpha\beta}\) the hopping amplitude from sublattice \(\beta\) in the reference cell to sublattice \(\alpha\) in the cell displaced by \(\mathbf R\).
The nonzero amplitudes are
\begin{align}
    \notag t_{\mathbf 0,11}&=A^2+1, & t_{\pm\hat{\mathbf y},11}&=A, \\[2pt]
    \notag t_{\mathbf 0,22}&=A^2+1, & t_{\pm\hat{\mathbf x},22}&=A, \\[2pt]
    t_{\mathbf 0,12}&=A^2, &
    t_{-\hat{\mathbf x},12}&=A, &
    t_{-\hat{\mathbf y},12}&=A, &
    t_{-(\hat{\mathbf x}+\hat{\mathbf y}),12}&=1.
    \label{eq:checkerboard_hops}
\end{align}
All other \(t_{\mathbf R,\alpha\beta}=0\).
Hopping amplitudes in opposite directions are identical, \(t_{-\mathbf R,\alpha\beta}=t_{\mathbf R,\alpha\beta}\).
These hoppings define the checkerboard-like lattice shown in Fig.~\ref{fig:generalized_checkerboard_lattice}.

We now consider normalized CLS (CLS normalized to 1) by multiplying a normalization constant \(1/\sqrt{2A^2+2}\).
The zeroth-order term of \(\alpha^2(\bk)\) becomes 1.
The overlaps of the CLS in the \(x\) and \(y\) directions are given by \(A/(2A^2+2)\).
Defining
\begin{align}
    A' \equiv \frac{2A}{2A^2+2}=\frac{A}{A^2+1},
\end{align}
we have
\begin{align}
    \alpha^2(\bk)=1+A'\big(\cos k_x+\cos k_y\big).
\end{align}
The direct gap occurs at \(\bk=(\pi,\pi)\) and is
\begin{align}
    \Delta = 1-2A'.
\end{align}

We expand near \(\bk=(\pi,\pi)\). Write
\begin{align}
    k_i=\pi+q_i \quad (i=x,y), \qquad |q_i|\ll 1,
\end{align}
so that \(\cos k_i=\cos(\pi+q_i)=-\cos q_i\simeq -\big(1-\tfrac{q_i^2}{2}\big)\).
Then
\begin{align}
    \alpha^2(\bk)
    &\simeq 1 + A'\!\left(-2 + \frac{q_x^2+q_y^2}{2}\right)
    = (1-2A') + \frac{A'}{2}(q_x^2+q_y^2).
\end{align}

With the BCLS choice
\(\phi_{\mathrm{BCLS}}(\bk)=\big(A+e^{ik_x},\;-\,[A+e^{-ik_y}]\big)^{\mathsf T}\),
the projector matrix elements (normalized by \(2(A^2+1)\)) expand as
\begin{widetext}
\begin{align}
    \notag
    \frac{|\phi_{\mathrm{BCLS}}(1,\bk)|^2}{2(A^2+1)}
    &\simeq \left(\frac{1}{2}-A'\right) + \frac{A'}{2}\,q_x^2,\\
    \notag
    \frac{|\phi_{\mathrm{BCLS}}(2,\bk)|^2}{2(A^2+1)}
    &\simeq \left(\frac{1}{2}-A'\right) + \frac{A'}{2}\,q_y^2,\\
    \label{eq:square_fb_almost_singular_fixed}
    \frac{\phi_{\mathrm{BCLS}}^*(1,\bk)\,\phi_{\mathrm{BCLS}}(2,\bk)}{2(A^2+1)}
    &\simeq \frac{1}{2A^2+2}\Big[
        -\,(A-1)^2
        - i(A-1)\,(q_x+q_y)
        - \frac{A-1}{2}\,(q_x^2+q_y^2)
        + q_x q_y
    \Big].
\end{align}
\end{widetext}
(Here we used \(e^{ik_x}=-e^{iq_x}\) and \(e^{-ik_y}=-e^{-iq_y}\) near \((\pi,\pi)\).)

\begin{figure}[t]
    \centering
    \includegraphics[width=\linewidth]{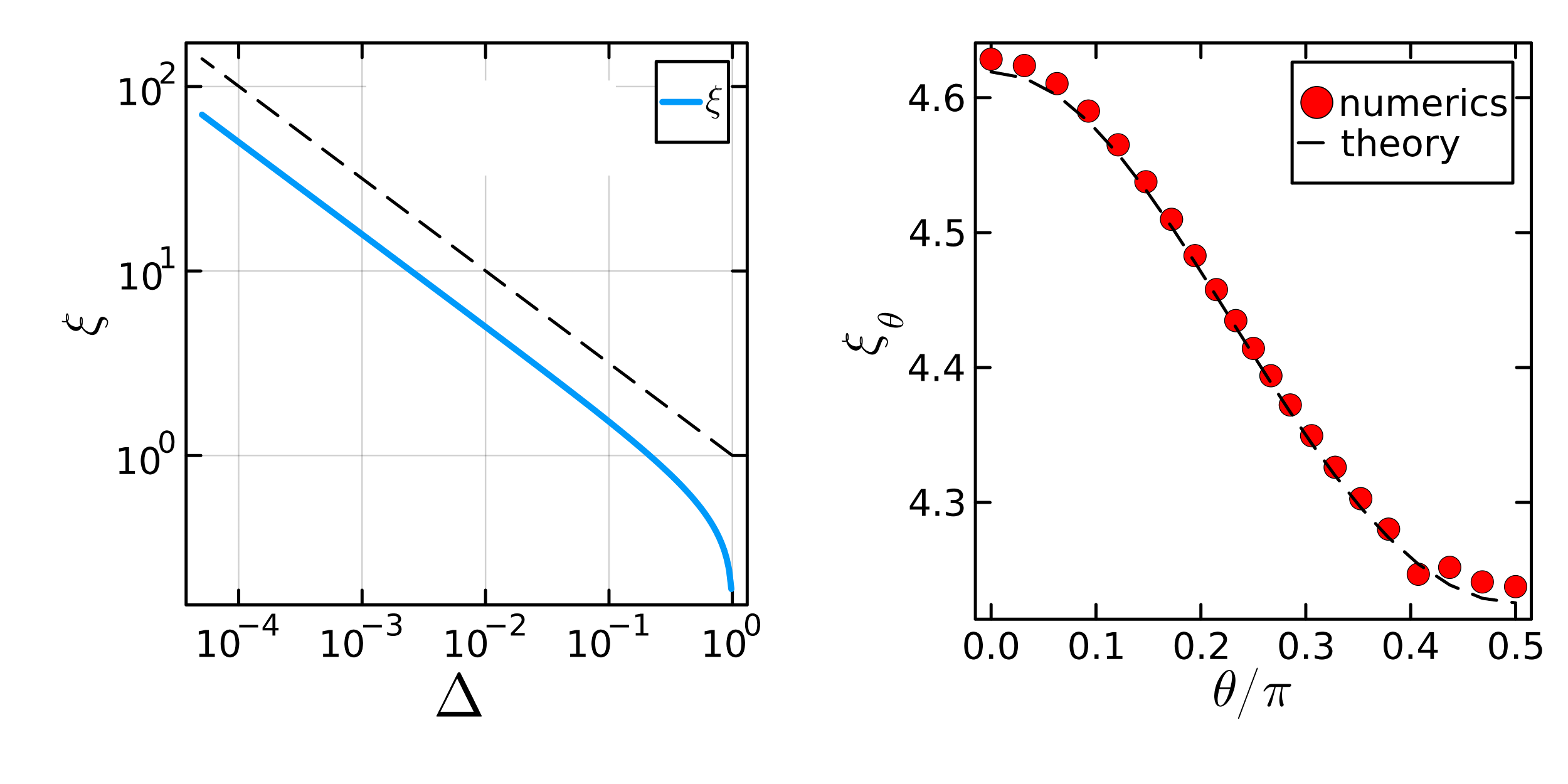}
    \caption{
        Dependence of the localization length \(\xi\) of the square-lattice flat band projector on parameters.
        (a) Scaling of \(\xi\) with decreasing band gap \(\Delta\).
        The dashed line shows the \(\Delta^{-1/2}\) trend.
        (b) Directional dependence of \(\xi\) versus the slice direction \(\theta=\arctan(y/x)\) in the 2D projector, where \((x,y)\) labels a lattice site.
    }
    \label{fig:endmatfig}
\end{figure}

Note that as \(A\to 1\) the constant and linear terms in
Eq.~\eqref{eq:square_fb_almost_singular_fixed} vanish while the quadratic terms remain.
For any sublattice choice, this yields the correlation length
\(\xi \propto (1-2A')^{-1/2}=\Delta^{-1/2}\) for small \(A'\), as expected.
The scaling and the directional dependence of \(\xi\) are shown in Fig.~\ref{fig:endmatfig}.

\subsection{Impossibility of singular flat bands in one dimension}

In one dimension, each component of the BCLS, \(\phi_{\mathrm{BCLS}}(\mu,k)\), is a Laurent polynomial in \(z = e^{ik}\), i.e. \(f_\mu(z) \in \mathbb{C}[z,z^{-1}]\). 
Assume there exists a momentum \(k_0\) (equivalently \(z_0 = e^{ik_0} \neq 0\)) such that \(f_\mu(z_0) = 0\) for all \(\mu\). 
Then each \(f_\mu(z)\) has a zero at \(z_0\), so it can be written as \(f_\mu(z) = (z - z_0)\,g_\mu(z)\) with \(g_\mu(z) \in \mathbb{C}[z,z^{-1}]\). 
Hence all components share the nontrivial common factor \((z - z_0)\). By definition, an irreducible BCLS does not allow any such common factor (other than overall constants or powers of \(z\), which are trivial). 
Therefore this condition cannot be satisfied in one dimension, and a singular flat band is impossible for \(d = 1\).

\subsection{Examples of nongeneric case}
\label{app:nongeneric_sfb}

We provide here some comments and examples for the case of a singular \(B\) in Eq.~\eqref{eq:pkq3}.
Let us first discuss how the matrix \(B\) is related to the flat band BCLS for the singular flat band.
For simplicity, let us consider real CLS.
In this case we have BCLS \(\vec\phi_\bcls(\bk_0)=0\) for real \(\bk_0\), and around this point we expand the BCLS as
\begin{align}
    \label{eq:bcls_expansion}
    \vec\phi_\bcls(\bk_0+\bq)=i\,b\,\bq+O(|\bq|^2),
\end{align}
where \(b\) is a \(u\times d\) real matrix (with \(u\) the number of bands/orbitals), and \(\bq=\bk-\bk_0\).
The matrix \(b\) encodes the directional dependence of the normalized Bloch flat band eigenstate~\cite{rhim2019classification}:
approaching \(\bk_0\) along direction \(q_i\) gives the vector proportional to the \(i\)th column of \(b\).
From Eq.~\eqref{eq:bcls_expansion} we find \(\alpha^2(\bk)=\vec\phi^\dagger_\bcls\vec\phi_\bcls\), so that
\begin{align}
    \alpha^2(\bk_0+\bq)=\bq^T (b^\dagger b)\,\bq + O(|\bq|^4).
\end{align}
We identify \(B=b^\dagger b\), a positive semidefinite matrix, ensuring \(\alpha^2(\bk)\geq 0\).

In the main text we assumed positive-definiteness of \(B\).
For a singular \(B\) we have \(B\bq_\parallel=0\) for at least one \(\bq_\parallel\).
There are several possibilities:
(i) \([\vec\phi(\bk)]_\mu\) are independent of \(\{\bq_{\parallel}\}\) and so is \(\alpha(\bk)^2\), and the band touching is not point-like.
Then the integral in Eq.~\eqref{eq:projector_fourier} is separable in \(\bq_\parallel\) and the flat band projector is compact in the directions corresponding to the nullspace of \(B\).
If the dimension of the null space of \(B\) is \(m\), then the exponent of the power-law spatial decay in the orthogonal directions is \(d-m\) and not \(d\), as in the generic case.
(ii) \(\bq_{\parallel}\)-related terms appear in higher orders of the expansion of \(\alpha(\bk)\), corresponding to flat bands with anisotropic touchings (e.g., mixed quadratic and quartic).
Handling this case requires a change of basis into \(\bq_{\parallel}\) and the rest \(\{\bq_{\perp}\}\).

We consider a simple 2D example falling under (ii), with a singular \(B\):
\begin{align}
    \label{eq:2d_bcls_example}
    \vec\phi_\bcls(\bk)=
    \begin{bmatrix}
        e^{ik_x}-1\\
        (e^{ik_y}-1)^2
    \end{bmatrix}.
\end{align}
Near the singularity \(\bk=0\), we have
\begin{align}
    \vec\phi_\bcls(\bk)\sim i
    \begin{bmatrix}
        1 & 0\\
        0 & 0
    \end{bmatrix}
    \begin{bmatrix}
        k_x\\ k_y
    \end{bmatrix}
    -
    \begin{bmatrix}
        -\frac{1}{2}k_x^2\\
        -k_y^2
    \end{bmatrix},
\end{align}
and
\begin{align}
    B =
    \begin{bmatrix}
        1 & 0 \\ 0 & 0
    \end{bmatrix}.
\end{align}
The asymptotic decay of the projector integral~\eqref{eq:projector_fourier} is governed by the Taylor expansion near the singularity:
\begin{gather}
    \notag P(\br)\sim \int_{\mathbb{R}^2} \mathcal P_{\mathrm{IR}}(k_x,k_y)\,e^{i(k_x x+k_y y)}\,dk_x\,dk_y,\\
    \mathcal P_{\mathrm{IR}}(k_x,k_y)=\frac{k_x^2}{k_x^2+k_y^4}.
\end{gather}
The integrand is a homogeneous function:
\begin{align}
    \mathcal P_{\mathrm{IR}}(k_x,k_y)=\mathcal P_{\mathrm{IR}}(\lambda k_x,\lambda^{1/2}k_y).
\end{align}
This implies the following scaling in real space:
\begin{align}
    P(x,y)=\lambda^{-3/2}P(\lambda x,\lambda^{1/2}y).
\end{align}
Hence \(P(0,y)\sim y^{-3}\) and \(P(x,0)\sim x^{-3/2}\), or in combined form
\begin{align}
    P(x,y)\sim \frac{1}{y^{3}+x^{3/2}}.
\end{align}

We also note that \(B=b^\dagger b\) is always singular if the number of bands \(u<d\) (\(d>2\)) in the time-reversal case, because \(b\) is \(u\times d\) and \(B\) is rank deficient.
The flat band projector then has an anisotropic decay.
As an example, consider a 3D BCLS with two orbitals
\begin{align}
    \label{eq:3d_2band_bcls}
    \vec\phi_\bcls(\bk)=
    \begin{bmatrix}
        e^{ik_x}+e^{ik_y}-2\\
        e^{ik_z}-1
    \end{bmatrix}.
\end{align}
The singular band touching is at \(\bk=0\).
Near this singularity, the BCLS is
\begin{align}
    \label{eq:3d_2band_bcls_taylor}
    \vec\phi_\bcls(\bk)\sim
    i\begin{bmatrix}
        1 & 1 & 0\\
        0 & 0 & 1
    \end{bmatrix}
    \begin{bmatrix}
        k_x \\ k_y \\ k_z
    \end{bmatrix}
    = b
    \begin{bmatrix}
        k_x \\ k_y \\ k_z
    \end{bmatrix}.
\end{align}
We have
\begin{align}
    B = b^\dagger b =
    \begin{bmatrix}
        1 & 1 & 0\\
        1 & 1 & 0\\
        0 & 0 & 1
    \end{bmatrix},
\end{align}
which is singular, with \(\bq_{\parallel}=[-1,1,0]^{\mathrm T}\).
Repeating the homogeneous-function argument, the decay is \(|\br|^{-5/2}\) in all directions except along \(\br=[-1,1,0]\), where we have \(|\br|^{-5}\).

Now let us allow CLS amplitudes to take imaginary values.
The matrix \(b\) may be complex, and the null space of \(b\) may then be spanned only by complex vectors, which are not relevant for real momenta \(\bk\in\mathbb{R}^d\).
A more accurate rank counting can be done by separating the real and imaginary parts of \(\vec \phi\).
If the number of bands is \(u\), the matrix \(b\) should be treated as \(2u\times d\) (over \(\mathbb{R}\)) instead of \(u\times d\) (over \(\mathbb{C}\)).
In other words, even if \(u<d\), it can happen that the decay is homogeneous.
A simple example is a slightly tweaked variant of Eq.~\eqref{eq:3d_2band_bcls} which breaks time-reversal symmetry:
\begin{align}
    \vec\phi_\bcls(\bk)=
    \begin{bmatrix}
        (e^{ik_x}-1) + i(e^{ik_y}-1)\\[4pt]
        e^{ik_z}-1
    \end{bmatrix}.
\end{align}
For this model, near \(\bk=0\) we have
\begin{align}
    \vec\phi_\bcls(\bk)\sim i
    \begin{bmatrix}
        1 & i & 0\\
        0 & 0 & 1
    \end{bmatrix}
    \begin{bmatrix}
        k_x\\ k_y\\ k_z
    \end{bmatrix}.
\end{align}
Here, the first-order term vanishes along the \((i,-1,0)\) direction, but for \(\bk\in\mathbb{R}^d\) the first-order term never vanishes.
Instead, one may consider the form with separate real and imaginary parts
\begin{align}
    \begin{bmatrix}
        \Re[\phi_\bcls(\bk)]_a\\
        \Im[\phi_\bcls(\bk)]_a\\
        \Re[\phi_\bcls(\bk)]_b\\
        \Im[\phi_\bcls(\bk)]_b
    \end{bmatrix} \sim
    \begin{bmatrix}
        0 & -1 & 0\\
        1 & 0 & 0\\
        0 & 0 & 0\\
        0 & 0 & 1
    \end{bmatrix}
    \begin{bmatrix}
        k_x\\ k_y\\ k_z
    \end{bmatrix},
\end{align}
which correctly counts the rank as 3, equal to the dimension \(d=3\).

\subsection{Anisotropic Lieb lattice}

We start with the BCLS
\begin{align}
    \label{eq:bcls_anisotropic_lieb}
    \phi_\bcls(\bk)=\begin{bmatrix}
        -1 - e^{ik_x}\\
        2 + e^{ik_y} + e^{-ik_y}
    \end{bmatrix}.
\end{align}
To construct the chiral flat band with the BCLS given in Eq.~\eqref{eq:bcls_anisotropic_lieb}, first one considers the vector \(v\) which satisfies \(v^\dagger \phi = 0\).
Then the chiral flat band is
\begin{align}
    \mathcal H(\bk)=\begin{bmatrix}
        0 & v^\dagger\\
        v & 0
    \end{bmatrix}=\begin{bmatrix}
        0 & 2 + e^{ik_y} + e^{-ik_y} & 1 + e^{ik_x}\\
        1 + e^{ik_x} & 0 & 0\\
        2 + e^{ik_y} + e^{-ik_y} & 0 & 0
    \end{bmatrix}.
\end{align}
This model possesses the BCLS with an additional entry of zero amplitude:
\begin{align}
    \label{eq:bcls_anisotropic_lieb_with_minor}
    \phi_\bcls(\bk)=\begin{bmatrix}
        0\\
        -1 - e^{ik_x}\\
        2 + e^{ik_y} + e^{-ik_y}
    \end{bmatrix}.
\end{align}
One can easily check that \(\mathcal H (\bk)\,\phi_\bcls(\bk)=0\), confirming this is a BCLS eigenstate of the central flat band at \(E=0\).
The lattice structure and the band structure are illustrated in Fig.~2(a) and (b).

\bibliography{ref}

\end{document}